# Molding the flow of light on the nanoscale: from vortex nanogears to phase-operated plasmonic machinery

**Svetlana V. Boriskina*[a] and Björn M. Reinhard*[b]**

Efficient delivery of light into nanoscale volumes by converting free photons into localized charge-density oscillations (surface plasmons) enables technological innovation in various fields from biosensing to photovoltaics and quantum computing. Conventional plasmonic nanostructures are designed as nanoscale analogs of radioantennas and waveguides. Here, we discuss an alternative approach for plasmonic nanocircuit engineering that is based on molding the optical powerflow through 'vortex nanogears' around a landscape of local phase singularities 'pinned' to plasmonic nanostructures. We show that coupling of several vortex nanogears into transmission-like structures results in dramatic optical effects, which can be explained by invoking a hydrodynamic analogy of the 'photon fluid'. The new concept of vortex nanogear transmissions (VNTs) provides new design principles for the development of complex multi-functional phase-operated photonics machinery and, therefore, generates unique opportunities for light generation, harvesting and processing on the nanoscale.

## Introduction

Noble metal nanoparticles efficiently absorb and re-radiate electromagnetic radiation in the visible and infra-red spectral range[1-3]. Incident light that is in resonance with the coherent collective electron oscillations of the noble metal's conduction band electrons can induce large amplitude oscillations, the so called localized surface plasmons (LSPs). The resulting alternating current causes an individual nanoparticle to act as a dipole antenna (Fig. 1a). Work by different groups has already demonstrated the applicability of radio-frequency (RF) antenna concepts to metallic nanostructures with a spectral response in the visible regime[4-11], resulting in successful demonstrations of optical analogs of Hertzian dipole antennas[6, 12], Yagi-Uda antennas[7, 13] and phased antenna arrays.[14, 15]

Nanolocalized high-intensity LSP fields can be used to greatly modify various radiative and non-radiative optical processes of individual molecules and materials (such as elastic and inelastic scattering, fluorescence, etc)[4, 16-20]. The strength of light-matter interactions depends on the optical field intensity, and thus the design of plasmonic nanoelements typically is centered on efficient nanoconcentration of light. However, it is important to recognize that the proper design of the field phase landscape plays a crucial role in shaping the spatial and spectral patterns of light intensity in the optical interference field. Optical antenna theory concepts are largely based on exploiting the effects of constructive interference between the fields emitted by individual nanoantennas. Prominent examples are periodic nanoantenna arrays, which serve as optical analogs of phased RF antenna arrays. In properly engineered array configurations, constructive interference of partial scattered fields leads to the formation of sharp resonant features in their scattering spectra, accompanied by strong resonant enhancement of the field intensity[14, 21-25] and of the local density of optical states (LDOS) near individual nanoantennas[15, 26, 27].

Another phase-related phenomenon used in the engineering of the scattering response of plasmonic nanostructures is the abrupt phase jump of the scattered field at the frequency of LSP resonance of a plasmonic nanoantenna driven by external optical field. In fact, this feature is manifested in the frequency spectrum of any resonant structure driven by an external field, with the classical forced oscillator being the simplest example. The phase of the oscillator (or that of the wave function in the case of the electromagnetic scattering problem) changes by π when the frequency is scanned through the resonance[28, 29]. This effect is a manifestation of the delay between the action of the driving force and the response of the oscillator (below the resonance the oscillator is in phase with the driving force, and it becomes out of phase above the resonance).

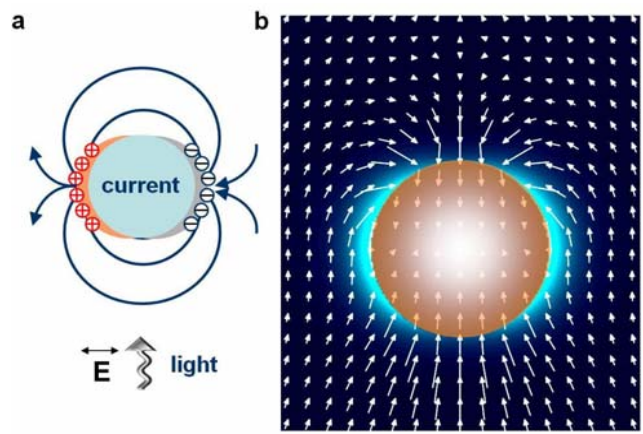

**Fig. 1** Plasmonic nanoantenna concept. (a) Collective oscillations of conduction electrons in metal nanoparticle excited by propagating light create regions of positive and negative surface charge, resulting in the generation of secondary dipole fields around the particle. (b) Optical powerflow around a noble-metal nanosphere at the frequency of its dipole LSP resonance shows strong deflection of incident light towards the particle. Hereafter, the arrows point into the direction of the local powerflow; the length of each arrow is proportional to the local value of the Poynting vector amplitude.

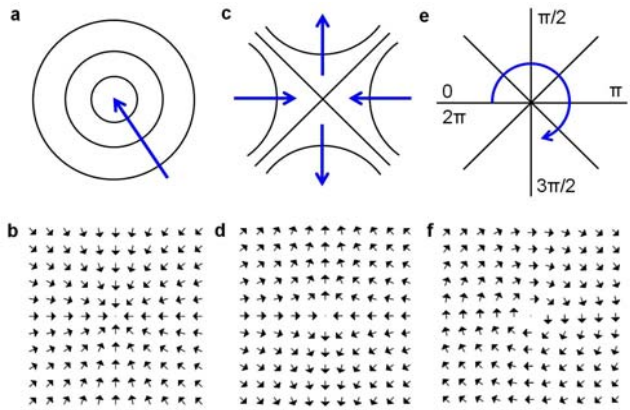

**Fig. 2** Phase singularities and stationary points of the optical field and corresponding topological features of the powerflow. (a) Phase maximum – a stationary point of optical field phase and (b) the corresponding powerflow sink. (c) Saddle point of the field phase and (d) the corresponding saddle point in the powerflow distribution. (e) Phase vortex – the point of undefined phase of optical field, where the whole 2π range of phase values co-exist. (f) Powerflow vortex (also referred to as a powerflow center) corresponding to the phase vortex in (e). The blue arrows indicate the direction of the phase increase in (a), (c) and (e).

The resonant phase change of a forced oscillator is a pre-cursor to engineering a Fano resonance[30] in the structure composed of two (or more) coupled oscillators.[31] In particular, coupled-oscillator structures can be designed such that the amplitude of the forced oscillator is suppressed at the eigenfrequency of the second oscillator as a result of destructive interference between the external force and the second oscillator. Fano-type plasmonic mode coupling[32-38] and a closely-related effect of optically-induced transparency[39, 40] have recently attracted a lot of interest in plasmonics. The possibility of generating sharp spectral features by coupling broad 'bright' plasmonic resonances to more narrow 'dark' modes (those that cannot be activated via the far-field excitation by propagating waves) promises many useful applications in bio(chemical)-sensing and optical communications. However, the sharp spectral dips generated in plasmonic nanostructures via the Fano effect typically do not overlap with the intensity peaks in the nanostructure near-field spectrum.[41]

The spatial phase landscape of the excitation field can also play an important role in manipulating light-nanostructure interactions. For example, temporal shaping of the phase and amplitude of ultrashort pulses can be used to control the spatial distribution of the light intensity in plasmonic nanostructures[42-44]. Alternatively, the spatial distribution of electromagnetic hot spots can be manipulated by continuous illumination of nanostructures with spatially phase-shaped fields (such as e.g. high-order Hermite-Gaussian and Laguerre-Gaussian beams)[45, 46]. The phase-operated spatial control of nano-optical fields offers many opportunities in microscopy, optical data storage and bio(chemical)-sensing.

The global spatial structure of the optical interference field depends dramatically on the fine structure of the field phase landscape, which is defined by the presence and spatial positions of the topological features, such as phase singularities and stationary nodes[47-49] (Fig. 2a,c,e). Phase singularities are the points of destructive interference in the optical field, where the field intensity vanishes and, as a result, the phase remains undefined (i.e. the whole 2π range of phase values co-exist). The simplest analogy is the undefined time zone of the north (south) pole. The stationary nodes, where the phase gradient of the electromagnetic field vanishes, include local extrema (maxima and minima) and saddle nodes. In two dimensional space, the phase singularities appear as isolated points, while in three dimensions they are curved lines, which exist along the intersection of the real and imaginary zero surfaces.

As optical energy flows in the direction of the phase change, the local powerflow is dramatically modified in the vicinity of phase singularities and stationary nodes. For example, the optical powerflow around a noble metal nanoparticle at its LSP resonance wavelength is shaped by the presence of saddle nodes in the particle's near-field (Fig. 1b). The on-resonance powerflow distribution shown in Fig. 1b illustrates an answer to the question 'How can a particle absorb more than the light incident on it?', famously posed and answered by C. Bohren[50]. In general, the electromagnetic powerflow is evaluated as the local time-averaged Poynting vector $\mathbf{S} = 1/2\,\mathrm{Re}[\mathbf{E}\times\mathbf{H}^*]$, where $\mathbf{E}$ and $\mathbf{H}$ are electric and magnetic field vectors, respectively. Accordingly, the topological features of the Poynting vector can be divided into two categories – those related to the phase singularities of the electromagnetic field and those associated with the field stationary phase nodes[50-53]. The stationary nodes include local extrema (maxima and minima) (Fig. 2a) and saddle nodes (Fig. 2b), with the former giving rise to the powerflow sinks or sources, respectively (Fig. 2b), and the latter corresponding to the saddle nodes of the powerflow (Fig. 2d). In turn, phase singularities (Fig. 2e) are always accompanied by the circulation of the optical energy, resulting in the formation of so-called optical vortices (centers of power flow)[47, 48, 50, 54-56] (Fig. 2f).

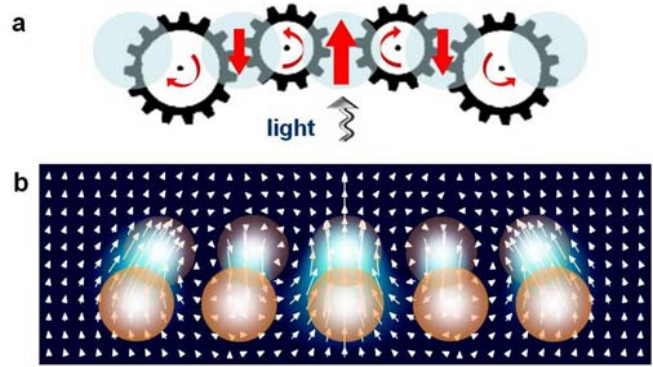

**Fig. 3** Vortex nanogear transmission concept. (a) VNT captures incoming light and twists it around phase singularities of the electromagnetic field 'pinned' to a plasmonic nanostructure. Coupled optical vortices - areas of circular motion of light flux – are pictured as gears made of light arranged in a transmission-like sequence. (b) Optical powerflow in the VNT composed of five noble-metal nanosphere dimers features several coupled optical vortices with alternating directions of rotation. The vector field of the powerflow is shown in the plane cutting through the centers of the dimers comprising the VNT.

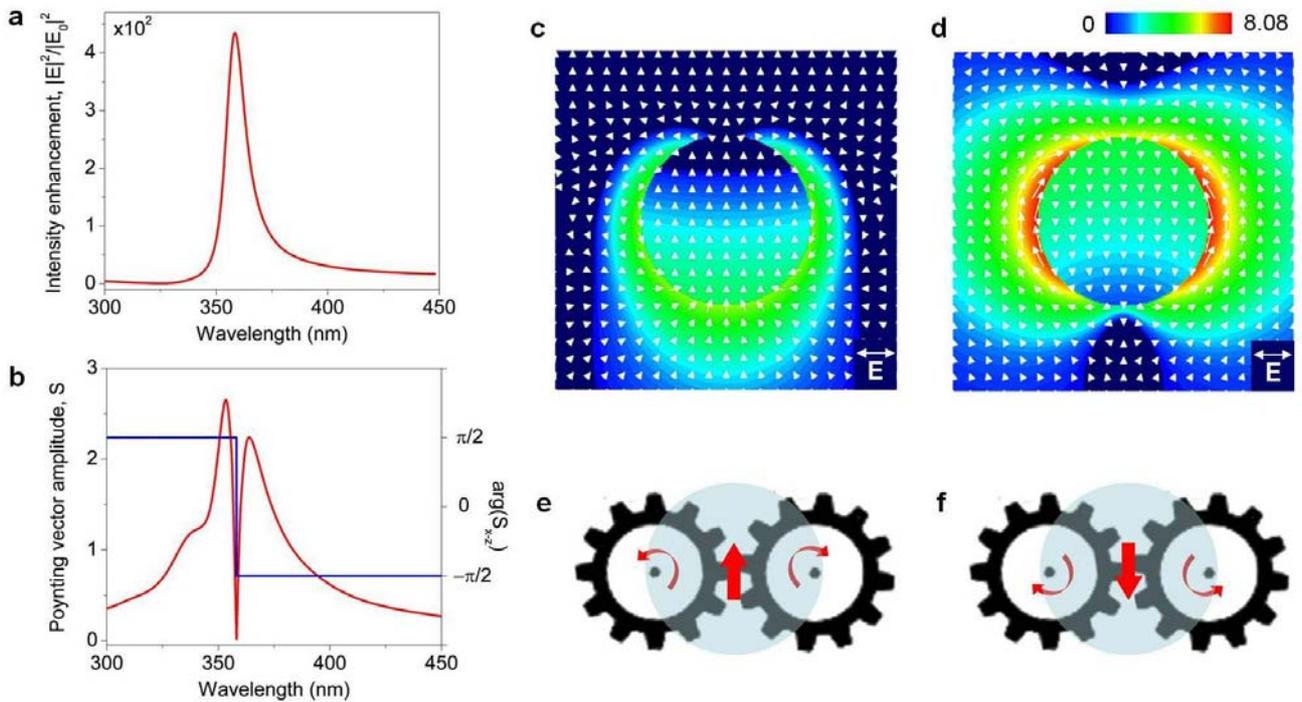

**Fig. 4** Formation of vortex nanogears around isolated nanoparticles. (a) Electric field intensity enhancement on the surface of a single Ag nanosphere of 15nm radius as a function of wavelength of the incident linearly-polarized plane wave. (b) The amplitude (red) and phase (blue) of the Poynting vector evaluated at the nanoparticle center. (c,d) Poynting vector intensity distribution and powerflow through the nanoparticle at the short-wavelength (c) and long-wavelength (d) maximum of the Poynting vector amplitude in (b). (e,f) Schematics of the VNTs generated at the two peak flow resonances. The light flux in each nanogear is looped through the particle volume.

Understanding the origins and exploiting wave effects associated with phase singularities have been proven to be of high relevance for many branches of physics, including hydrodynamics, acoustics, quantum physics, and singular optics[47, 57, 58]. In nano-photonics and plasmonics, visualization of the powerflow has already been used to gain a deeper physical insight into the performance of electromagnetic elements[47-50, 55]. For example, observation of the complex spatial phase signatures provided the explanation of the origin of enhanced/suppressed transmission through sub-wavelength slits,[53] sub-wavelength focusing in superlenses,[59-61] scattering-induced enhancement/suppression of photocurrent in photodiodes[62] and nanoparticle-mediated nanoholes processing in dielectric substrates[63]. However, a systematic plasmonic nanocircuit design based on the manipulation of the fine structure of the optical energy flow on the nanoscale has not been attempted until recently.

In the following, we summarize a rational nanocircuit design strategy based on creation and deterministic positioning of phase singularities 'pinned' to plasmonic and optoplasmonic nanostructures. We show that this approach provides an alternative way of tailoring the spatial structure of the plasmonic near-field. Differently from the traditional approaches based on engineering constructive interference of dipole fields emitted by individual nanoantennas, the outlined design strategy is based on engineering destructive interference of near-optical fields at pre-defined positions. This results in formation of phase singularities, which are accompanied by the circulating nanoscale powerflow (optical vortices). We discuss previously explored and novel plasmonic nanostructures whose unique optical properties can be explained by interactions of coupled nanoscale optical vortices. Through electromagnetic calculations, we reveal that the powerflow patterns in these nanostructures resemble multiple-gear transmissions (as schematically illustrated Fig. 3a), and term them 'vortex nanogear transmissions' (VNTs). VNTs are composed of coupled vortex nanogears made of light, with each nanogear rotating around an axis created by a field phase singularity. One example of the powerflow distribution in a possible VNT configuration to be discussed in the following is shown in Fig. 3b. Finally, we demonstrate that the new design approach, which is centered around molding the flow of light, can benefit greatly on drawing the analogies with hydrodynamics, which studies the flow of fluids. Applications of the VNT concept to the design of nanoantennas, bio(chemical) nanosensors, nanoscale transmission lines, and functional elements of plasmonic nanocircuits are also discussed.

## Computation approach

To get physical insight into the intricate features of the local light flow through plasmonic nanostructures and to use this insight to design vortex nanogear transmissions, we make use of rigorous semi-analytical method based on the generalized classical multi-particle Mie theory (GMT)[25, 51, 64-66]. The interparticle separations in all of the VNT structures to be discussed in the following are in the range of 1-50 nm, making them amenable to study in the frame of the classical electromagnetic theory[67]. GMT is a spectral method, which relies on the expansion of partial fields scattered by each particle into a series of functions that form a complete basis and on solving the truncated matrix equation to find the unknown expansion coefficients. In particular, if vector spherical harmonics centered around individual particles are used as the basis functions: $\mathbf{E}_{sc}^l = \sum_{(n)} \sum_{(m)} (a_{mn}^l \mathbf{N}_{mn} + b_{mn}^l \mathbf{M}_{mn})$, $l = 1,...L$, GMT provides an analytical solution for an arbitrary agglomerate of $L$ spherical particles. A general matrix equation for the Lorenz-Mie multipole scattering coefficients $a_{mn}^l$ and $b_{mn}^l$ is obtained by imposing the electromagnetic boundary conditions for the electric and magnetic fields. Differently from simpler asymptotic approaches, such as e.g., coupled dipole methods, in GMT calculations the contribution of a desired multipolar order can be taken into account by a proper truncation of the matrix equation. The contribution from the higher-order spherical harmonics (different from the dipolar one) becomes extremely important if nanoparticles are strongly-coupled to each other through their near-fields[68] and/or if they are large enough to support the higher-order resonances[51].

Once the system matrix equation is solved, the scattering, extinction and absorption cross-sections as well as the near-field intensity enhancement can be computed to any pre-defined level of accuracy. We calculate the volume-equivalent scattering and absorption efficiencies as the ratios of the scattering ($C_{sc}$) and absorption ($C_{abs}$) cross-sections of a nanostructure under plane-wave illumination to the geometric cross section of a volume-equivalent single sphere: $Q_{sc} = C_{sc}/\pi a_c^2$, $Q_{abs} = C_{abs}/\pi a_c^2$ where $a_c = (3V_c/4\pi)^{1/3}$, and $V_c$ is the combined volume of all the nanoparticles[25, 65]. We also used GMT to calculate the radiative rates of dipole emitters embedded in plasmonic nanostructures. In particular, the radiative decay rate $\gamma_r$ of the dipole $\mathbf{p}$ at the emission wavelength $\lambda_{em}$ is calculated as the power fraction radiated into the far field by integrating the energy flux through the closed surface surrounding both the dipole and the nanostructure. Experimentally obtained Ag and Au refractive index values from Johnson and Christy[69] are used in all the simulations.

## Powerflow around noble-metal nanoparticles

The powerflow around a noble metal nanoparticle shown in Fig 1b, which converges toward the nanoparticle and circulates between the **E** and **H** planes[50, 51], is a well-known characteristic feature of the nanoparticle dipole LSP resonance. A less known phenomenon is formation of pairs of optical vortices around noble-metal nanoparticles at the frequencies above and below their LSP resonance.[70] To illustrate this effect, in Fig. 4 we show the frequency spectrum of the light intensity generated on a 30nm Ag nanosphere illuminated by a linearly-polarized plane wave (Fig. 4a) and compare it to the corresponding spectra of the amplitude and phase of the local Poynting vector (Fig. 4b). The Poynting vector $\mathbf{S} = \{S_x, S_y, S_z\}$ is a 3D real-valued vector field. To visualize the optical powerflow in each coordinate plane, the phase of this vector field is defined as $\arg(S_{i-j}) = \arctan S_j/S_i$, $i, j = x, y, z$. The excitation of the dipole LSP resonance in the nanoparticle is marked by the pronounced peak in the field intensity spectrum shown in Fig. 4a. Fig. 4d demonstrates that at this point, the direction of the powerflow through the particle changes abruptly from forward ($\arg(S_{x-z}) = \arg(S_{y-z}) = \pi/2$) to backward ($\arg(S_{x-z}) = \arg(S_{y-z}) = -\pi/2$). Note that at the LSP resonance frequency all the flow lines converge towards the particle, and the backward flow behind the particle is governed by the presence of a saddle point in its shadow region (see Fig. 1a).[50-52]

The LSP resonance peak in the field intensity spectrum in Fig. 4a corresponds to the dip in the spectrum of the Poynting vector amplitude (power flux density) accompanied by the abrupt powerflow reversal (Fig. 4b). At the same time, two maxima can clearly be observed in the power density spectrum (Fig. 4b), which flank the LSP resonance dip on the longer- and shorter-wavelength sides. The spatial maps of the density and direction of the local optical power flux at the two peak wavelengths are plotted in Figs. 4c and 4d. These plots reveal the formation of pairs of coupled counter-rotating nanoscale optical vortices around the nanoparticle, constituting the simplest example of the two-gear vortex nanogear transmission (shown schematically in Figs. 4e,f). At the shorter-wavelength peak, the optical energy flows forward through the nanoparticle volume, and then backward on both sides of the particle to re-enter the particle from its illuminated side. At the longer-wavelength peak, the flow direction is reversed: the powerflow splits sideways before impinging on the particle and enters the particle volume through the shadow side of the particle.

Figure 5 offers a deeper insight into the intricate spatial phase structure of the optical field at and around the resonant frequency of the nanopartuicle dipole LSP mode. In particular, it demonstrates the migration of the powerflow topological features (vortices and saddle points) with the change of the wavelength, which is scanned through the LSP resonance. It can be seen that the on-resonance powerflow reversal observed in Fig. 4b occurs when the saddle node drifts through the position of the numerical detector at the center of the particle[51, 52]. This drift, accompanied by creation and annihilation of additional topological features (in accordance with the topological charge conservation principle[53, 71]) causes formation of oppositely-rotating coupled vortex nanogears around the nanoparticle at the frequencies above and below the LSP resonance.[70]

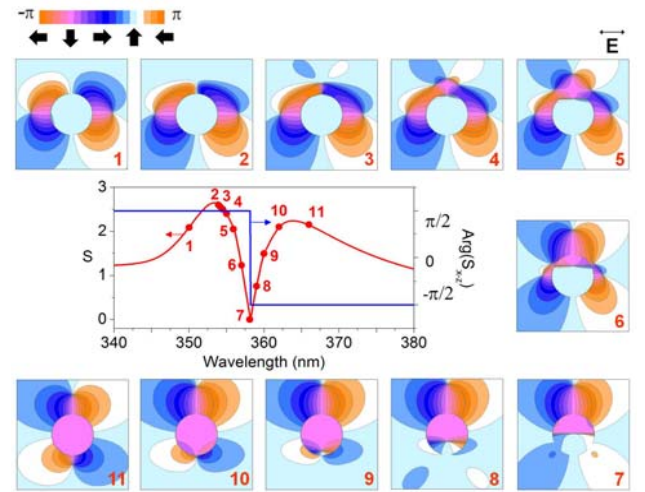

**Fig. 5** Evolution of the phase portrait of the Poynting vector around the dipole LSP resonance of a 30m Ag nanosphere. The phase distributions are plotted at the wavelengths marked with solid circles in the Poynting vector spectrum. Panels 1-2 and 10-11 feature counter-rotating vortex pairs accompanied by saddle nodes, which define the optical powerflow shown in Figs. 4c and 4d, respectively. Panel 7 corresponds to the nanoparticle LSP resonance and features a pair of saddle nodes that re-route the optical energy from the 'shadow' region back into the particle volume (Fig. 1b). The inset shows the powerflow direction corresponding to the color scheme of the Poynting vector phase pattern.

As the energy is re-circulated through the particle volume, the vortex nanogears forming around isolated nanoparticles can be classified as 'internal'. However, formation of internal nanogears in individual nanoparticles does not translate into efficient field enhancement (compare Figs. 2a and 2b). Furthermore, formation of complex flow patterns around individual particles often requires unusual material properties to reduce dissipative losses[51, 52]. Nevertheless, we demonstrate in the following that, unlike internal vortex nanogears induced in individual particles, rationally designed VNTs can sustain dramatic electromagnetic effects. In the frame of the proposed new approach, elemental building blocks of plasmonic nanocircuits are to be designed to feature phase singularities (resulting in the creation of vortex nanogears) at pre-defined positions. As the example in Fig. 4 reveals, the individual blocks may not exhibit any interesting electromagnetic behavior, and as such, they would have been discarded in the frame of the traditional design approaches.

Tailored coupling of individual vortex nanogears into nanogear transmissions, however, can induce the emergence of collective optical effects that create high field intensities and ultra-narrow resonance linewidths.

## The plasmonic nanolens as an internal VNT

First, we revisit a Stockman nanolens[72] – a self-similar chain of Ag nanospheres (Fig. 6a) – which produces resonant cascaded field enhancement in the focal point between the smallest spheres. We consider a three-sphere Ag nanolens of the same configuration as proposed in the pioneering Li *et al* paper[72] (the ratio of the consecutive radii is $r_{i+1}/r_i = 1/3$ and the distance between the surfaces of the consecutive nanospheres is $d_{i,i+1} = 0.6 r_{i+1}$). The structure is illuminated by a plane wave polarized along the nanolens axis. The spatial distributions of the nanolens near-field intensity shown in Figs. 6b and 6c are calculated at the wavelength of the near-field intensity resonance (b) and at an arbitrary chosen off-resonance wavelength (c). These plots reveal very similar spatial patterns, however, the focal-point intensity is about an order of magnitude higher at the resonance wavelength.

The conventional explanation of this effect is that larger nanoparticles have higher capacity of capturing light (characterized by their larger scattering cross-sections). The captured energy is then funneled from the large nanosphere along the nanolens via near-field dipole coupling to the smaller one, which in turn enables shrinking of the hop-spot size. However, a comparison of the far-field scattering spectrum (Fig. 6d) and the near-field intensity spectrum evaluated in the nanolens focal point (Fig. 6e) does not support this hypothesis. In particular, the focal-point intensity peak observed in Fig. 6e corresponds to a dip rather than a peak in the scattering efficiency spectrum. This dip is significantly blue-shifted from the broad scattering peak corresponding to the excitation of the dipole LSP mode in the largest nanosphere.

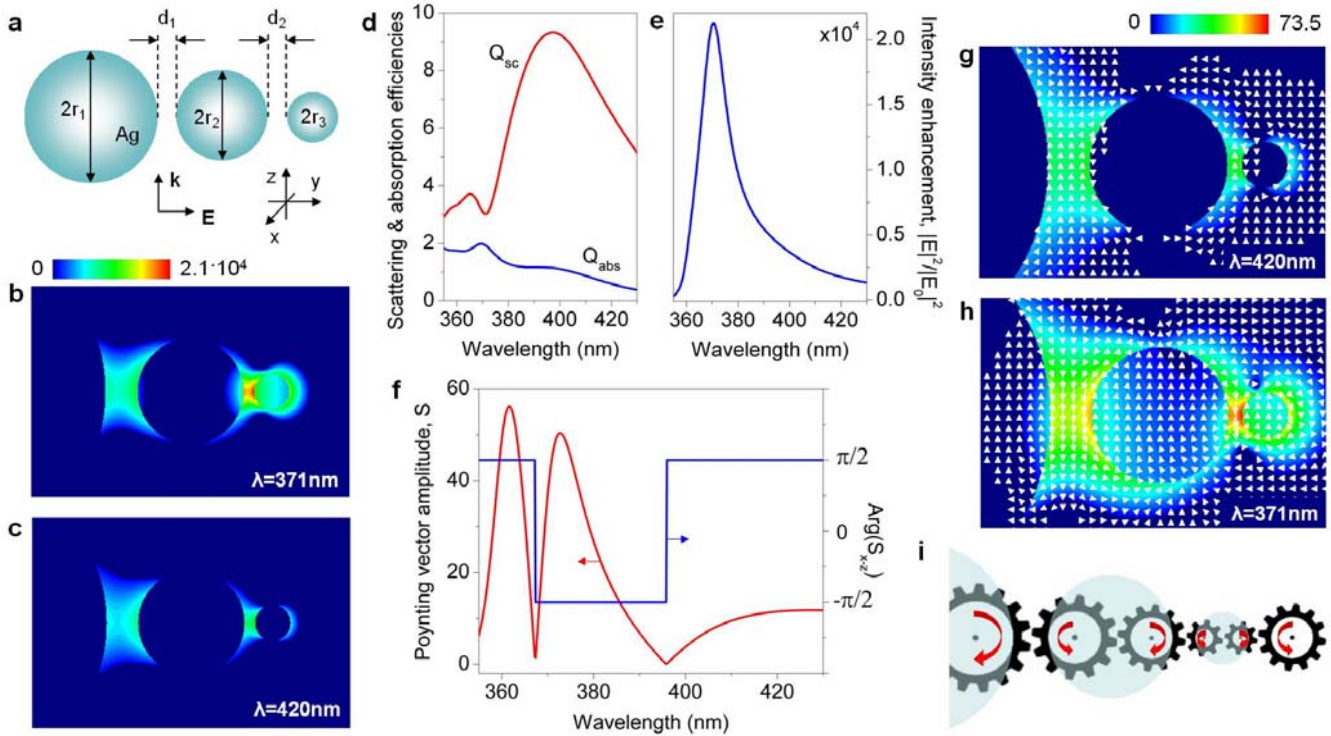

**Fig. 6** Plasmonic nanolens as an internal vortex nanogear transmission. (a) Schematic of the self-similar Ag nanolens proposed in Ref. 26 ($r_1$=45nm, $r_2$=15nm, $r_3$=5nm, $d_1$=9nm, $d_2$=3nm, ambient index $n$=1.0). (b,c) Electric field intensity distribution in the nanolens illuminated on- (b) and off-resonance (c) with the near-field intensity maximum of the nanolens. Far-field (d) and near-field intensity enhancement (e) spectra of the nanolens. (f) The amplitude of the Poynting vector and the phase of the Poynting vector in the x-z plane at the center of the nanolens narrower interparticle gap as a function of wavelength. (g,h) Poynting vector intensity distribution and powerflow around the nanolens off (g) and on (h) the peak intensity wavelength. (i) Schematic of the VNT generated in the nanolens at the peak intensity resonance. Light flux in each nanogear is looped through nanoparticles.

To understand the physical effects underlying the nanolens focusing and intensity enhancement mechanism, in Fig. 6f we plot the frequency spectra of the amplitude and phase of the Poynting vector evaluated at the center of the nanolens focal point (a gap between the smaller spheres). Figure 6f reveals a complex picture of powerflow through the nanolens focal point, with a reversal of the flow direction (from forward to backward) at the frequency of the near-field intensity resonance, followed by another reversal (from backward to forward) at the frequency of the scattering efficiency resonance. Vector fields of the powerflow through the nanolens off and on the near-field intensity resonance (Figs. 6g and 6h) illustrate the mechanism governing the light nanoconcentration in the nanolens. At the off-resonance wavelength, the optical power flows through the nanolens in the forward direction. At the near-field intensity peak wavelength, vortex nanogears form inside and near to the nanoparticles (Fig. 6h). Light circulation through these nanogears causes increased backward powerflow through the inter-particle gaps and results in the dramatic local field enhancement in the nanolens focal point. The coupled-vortex spatial picture resembles a complex gearbox (shown schematically in Fig. 6i) composed of gears of varying size, where the strongest electromagnetic hot-spot is created at the junction between the faster-turning smaller gears.

The visualization of the optical powerflow helps to reveal the importance of the nanospheres size mismatch for triggering the nanolensing effect. In individual nanoparticles, the far-field scattering resonance practically overlaps with the near-field intensity peak. However, the spectral position of the dipole LSP resonance of a smaller nanosphere is blue-shifted from that of a larger particle. As a result, the smaller particle LSP resonance falls into the spectral region characterized by the formation of a pair of vortex nanogears around the larger particle shown in Figs. 4c and 4e. The enhanced backward powerflow around the larger particle drives the circulating powerflow through the smaller one, resulting in the increase of the local field intensity in the nanolens focal point.

This vortex-driven light focusing mechanism does not necessarily require a self-similarity of the plasmonic structure (as was already concluded in the original Li *et al* paper[72] which started with the premise of the importance of the self-similarity). In fact, the main details of the VNT-driven focusing mechanism can be elucidated by studying a simple two-particle nanolens as shown in Fig. 7. Here, the radius of the larger nanoparticle is set at 40nm, and that of the smaller one (termed satellite) is varied from 40nm to 5nm. As can be seen in Fig. 7a, the intensity in the focal point (gap) of the nanolens increases with the decrease of the satellite size, while the spectral position of the intensity peak progressively blue-shifts. The frequency spectra of the powerflow direction through the gap (Fig. 7b) reveal that the observed field enhancement is related to the enhanced *backward* powerflow through the gap of size-mismatched dimers. Note that the symmetrical dimer is characterized by the *forward* powerflow through the gap (grey line). This reversed powerflow is driven by the formation of the internal nanogears on the shorter-wavelength side of the LSP resonance of the larger particle (see Figs. 4c,e). Further decrease of the satellite size blue-shifts the resonant peak into the frequency range where the particles have high dissipative losses. As a result, the near-field intensity reaches the highest value at the satellite radius of 14nm, but remains larger than that in a symmetrical dimer even at the satellite radius of 5nm.

## Optoplasmonic vortex nanogates

Strongly localized fields driven by the internal vortex nanogears enhance interactions between the light and the medium in the focal point of the plasmonic nanolenses. However, efficient on-chip light routing and transfer via *internal* VNTs is hampered by the dissipative losses caused by light re-circulation through the metal volume of nanoparticles. For many applications it is desirable to design plasmonic nanostructures capable of re-circulating light *outside* of the nanoparticle volume. This would increase the effective volume of light interaction with the host medium – which might be useful for sensing, photovoltaics, and active dynamic control of light in nanostructures – and would reduce both radiative and material dissipative losses.

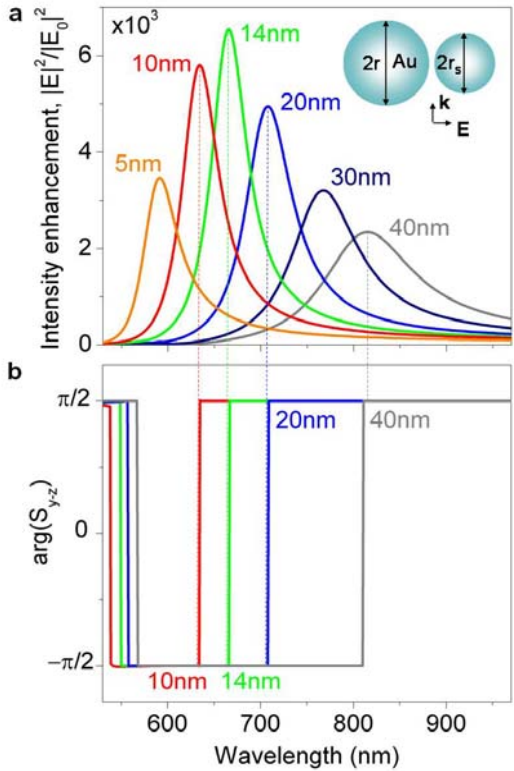

**Fig. 7** Double-particle plasmonic nanolens. (a) Electric field intensity enhancement in a dimer composed of size-mismatched Au nanoparticles (*r*=40nm, gap width *d*=1nm, ambient index *n*=1.44) as a function of the wavelength and the satellite radius $r_s$ (shown as labels) under illumination by a plane wave. Schematic of the nanolens geometry is shown in the inset. (b) The phase of the Poynting vector in the nanolens gap (dash lines mark the direction of the powerflow through the gap at select resonances).

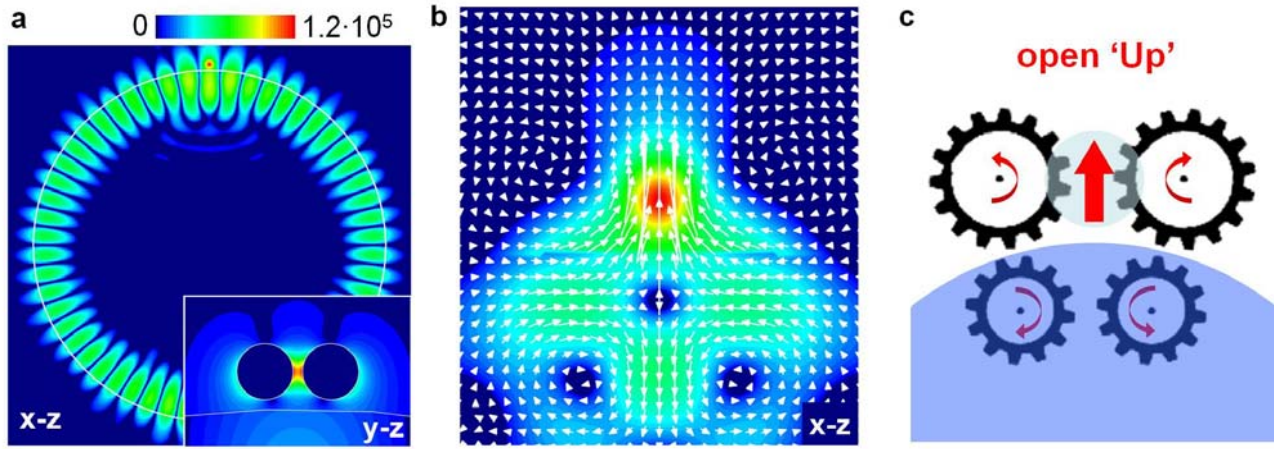

**Fig. 8** Optoplasmonic vortex nanogate. (a) E-field intensity distribution in the optoplasmonic structure composed of an Au nanoparticle dimer (nanosphere diameters 150nm, dimer gap 25nm) and a 5.6μm – diameter polystyrene microsphere. The structure is excited by an y-polarized plane wave at the wavelength of its $TE_{27,1}$ photonic-plasmonic resonance. The inset shows the on-resonance intensity distribution in the dimer gap. (b) Poynting vector intensity |S| maps and optical powerflow through the dimer gap (in the x-z plane at y=0) at the frequency of the photonic-plasmonic mode shown in (a). (c) Schematic of the vortex-operated nanogate in the 'open Up' position corresponding to the powerflow pattern in (b). Adapted from[65].

Fortunately, optical vortices and saddle nodes are characterized by the zero energy absorption ($\nabla \cdot \mathbf{S}(\mathbf{r}) = 0$), and thus can be created in the non-absorbing medium outside of particles. We have recently demonstrated that optical vortices can be generated in plasmonic nanostructures that are electromagnetically coupled to optical microcavities[73]. Dielectric and semiconductor microcavities – microspheres, microdisks, microtoroids, etc. – can trap and store incoming light by re-circulating it inside the cavity volume in the form of so-called whispering gallery (WG) modes[74-76]. WG modes are higher-azimuthal-order solutions of the Maxwell equations, and their electromagnetic wave functions are characterized by the $\exp\{im\varphi\}$, $m = 1,2,...$ phase dependence.[29, 77] This phase dependence is a clear signature of the presence of phase singularities in their electromagnetic fields. If the microcavity is excited at the wavelength of its WG mode, powerflow vortices form around the phase singularities in the resonant interference field inside and around the microcavity. These vortices can be threaded through plasmonic nanostructures immersed into the WG mode evanescent tail outside of the microcavity,[73] yielding strong resonant field enhancement within the nanostructure.

This effect is illustrated in Fig. 8, which shows a simple optoplasmonic structure composed of an Au nanodimer coupled to a polystyrene microsphere[73]. Under illumination by a plane wave, this optoplasmonic structure sustains hybrid photonic-plasmonic modes, such as the one shown in Fig. 8a. This mode results from the coupling between the high-Q $TE_{27,1}$ (with one radial and 27 azimuthal field variations) WG mode of the microsphere and a dipole LSP mode of the Au nanodimer, and is characterized by strong cascaded light enhancement in the nanodimer gap (see inset to Fig. 8a). This cascaded light enhancement is driven by the formation of coupled optical vortices inside the microcavity and around the nanodimer, which generate the enhanced powerflow through the gap as shown in Fig. 8b. By scanning the excitation wavelength through the hybrid photonic-plasmonic resonance, oppositely-rotating vortices can be forced to either approach and annihilate, or to nucleate as a pair. As a result, this optoplasmonic structure operates as a photonic nanogate, which directs the optical powerflow through the nanosized channel (dimer gap). The gate can be open either 'Up' (as shown schematically in Fig. 8c) or 'Down', corresponding to either forward or backward powerflow through the dimer gap. The gate can also be 'Closed' at select wavelengths where the Poynting vector amplitude vanishes resulting in no powerflow through the gap. The optoplasmonic vortex nanogate shown in Fig. 8 is an example of an *external* VNT structure, which circulates the optical energy through the ambient medium in the dimer gap rather than through the nanoparticles volumes. This reduces dissipative losses and translates into narrow resonant features, which might be useful for many applications in optical sensing and quantum information processing [73, 78, 79].

## External plasmonic nanogear transmissions

The concept of optoplasmonic vortex nanogates can be expanded to design more complex phase-operated structures[73]. However, realization of fully-plasmonic *external* VNTs would open tremendous opportunities for the low-loss light transfer and switching within nanoscale-size structures. The uniqueness of LSP resonances of noble-metal particles is in the extreme sub-wavelength localization of light they provide, which makes possible vortex engineering on the nanoscale. In Fig. 9a, we show an example of a fully-plasmonic nanostructure, which is designed to trap and recirculate the incoming light through a sequence of coupled vortices formed in the ambient medium in its near field. It consists of identical Ag nanosphere dimers arranged into a linear chain of varying length. The far- and near-field response of this nanostructure is studied in Fig. 9 under the normal illumination by a plane wave polarized along the dimers axes.

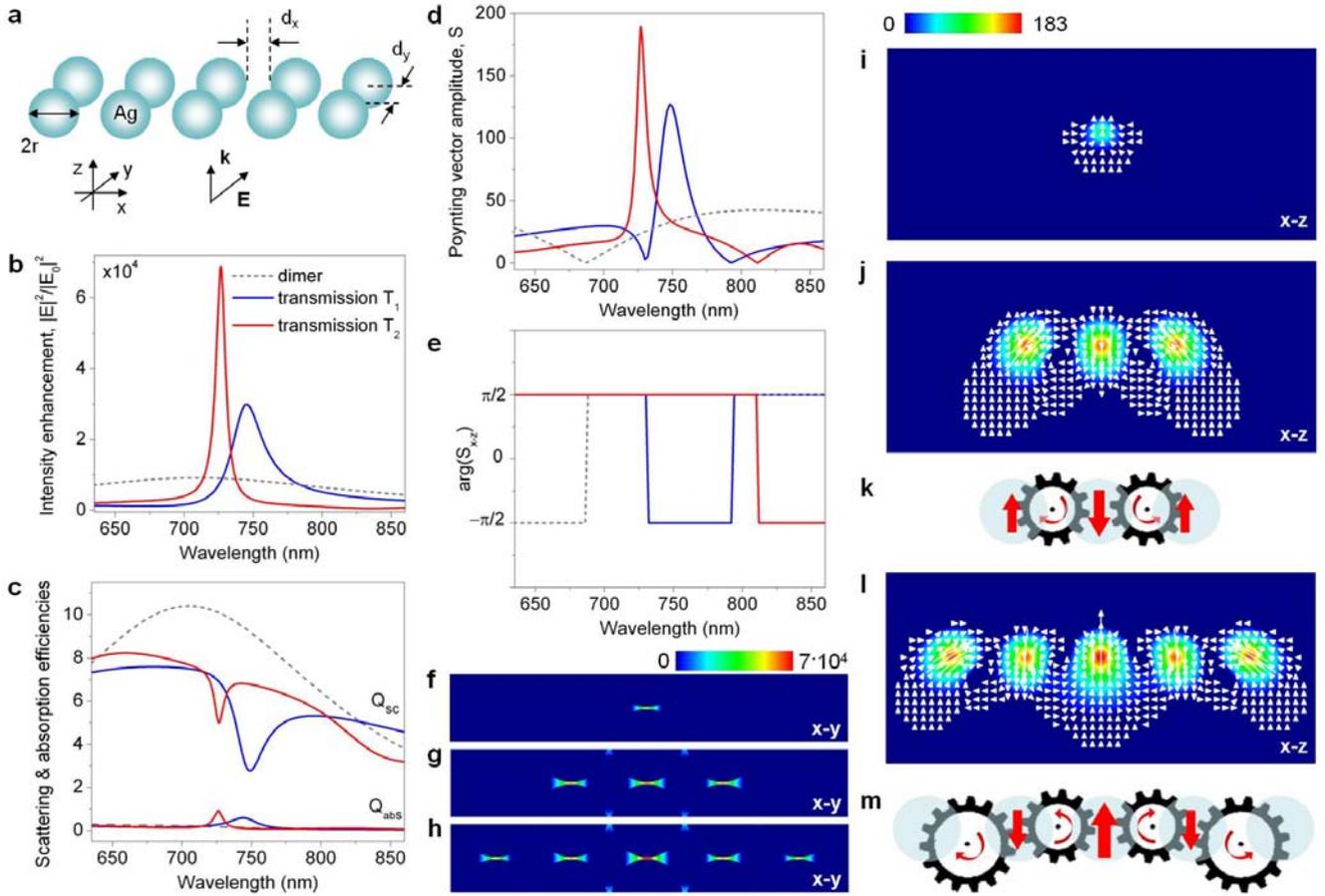

**Fig. 9** External plasmonic vortex nanogear transmissions. (a) Schematic of a linear chain of Ag nanoparticle dimers ($r$=50nm, $d_x$=10nm, $d_y$=3nm, ambient index $n$=1.33). Wavelength spectra of the electric field intensity enhancement in the gap of the central dimer (b), the far-field scattering and absorption efficiencies (c), the Poynting vector amplitude (d) and the phase of the Poynting vector in the x-z plane (e) in the gap of the central dimer of the nanostructures composed of three (solid blue lines) and five (solid red lines) nanodimers. The corresponding spectra of a single dimer are shown as dashed grey lines. The structures are illuminated by a plane wave with the electric field polarized along the dimers axes. Electric field intensity distribution in the gaps of a single dimer (f), three-dimer chain $T_1$ (g), and five-dimer chain $T_2$ (h). Poynting vector intensity distribution and powerflow through the gaps of the dimer (i), $T_1$ (j) and $T_2$ (l). Schematics of the VNTs generated in $T_1$ (k) and $T_2$ (m) at their resonant wavelengths. Light flux in each nanogear is looped between the particles.

Figures 9b and 9c compare the far-field scattering response and the near-field intensity spectrum of a conventional dimer-gap antenna[6, 8, 92] (i.e., an isolated dimer identical to a single unit in the chain shown in Fig. 9a) with those of linear-chain nanostructures composed of three (transmission $T_1$) and five (transmission $T_2$) dimers. In contrast to the frequency spectra of an isolated nanodimer, which feature broad overlapping scattering and near-field intensity bands, the near- and far-field spectra of $T_1$ and $T_2$ contain well-defined features. In particular, both structures exhibit sharp near-field intensity maxima (Fig. 9b), which are accompanied by sharp dips in their scattering spectra (Fig. 9c). The observed resonant drop in the scattering efficiency indicates their capability of trapping incoming light.

The increase of the dimer-chain VNT length results in the narrowing of the resonant spectral features, which is a signature of the increase of the dephasing time of the trapped light. The long dephasing time of the mode (characterized by a high quality (Q) factor) translates into its enhanced ability to accumulate the energy from the external excitation field. The resulting local field enhancement in the nanostructure is proportional to the Q value of the trapped-mode VNT resonance[80]. The near-field intensity spectra shown in Fig. 9b indeed demonstrate progressive severalfold increase of the field intensity generated at the centers of the longer VNTs, resulting from the increased Q-factors of their trapped-mode resonances.

Calculations of the Poynting vector in the VNT center show that the intensity build-up in the VNT interstitial spaces is driven by the strong resonant enhancement of the optical powerflow through the dimer gaps (see Fig. 9d). Furthermore, it can be seen that the direction of the flow through the VNT center reverses abruptly with the change of the wavelength of incident light (Fig. 9e). In particular, VNT $T_1$ is characterized by the resonantly-enhanced backward powerflow through its central gap, while VNT $T_2$ features strong forward powerflow at the resonant wavelength. Note that this complex picture of the powerflow cannot be extracted from the near-field intensity plots in the three nanostructures at their corresponding peak wavelengths (shown in Figs. 9f-h).

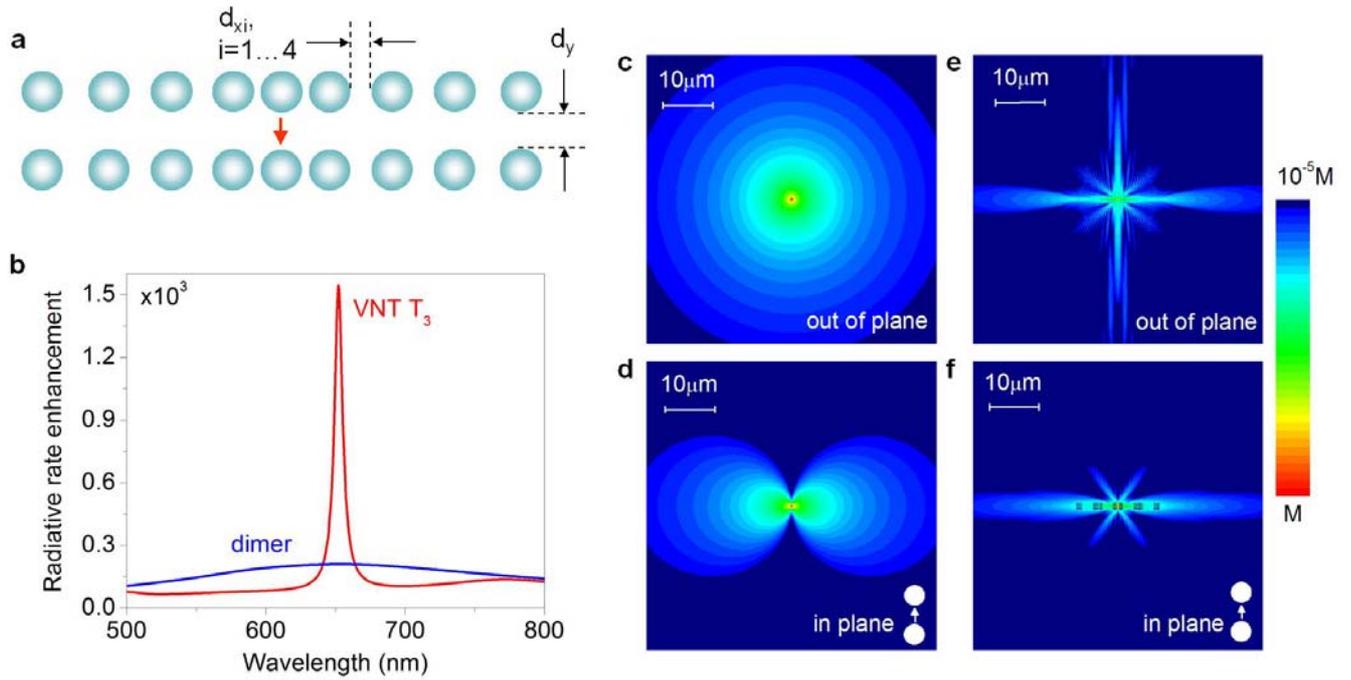

**Fig. 10** Radiative rate and emission patterns shaping of a dipole emitter via VNT mechanism. (a) Schematic of a dimer-chain VNT $T_3$ with a chirped period composed of 50-nm radius Ag nanosphere dimers (ambient index $n=1.33$; $d_{x1}=50$nm, $d_{x2}=100$nm, $d_{x3}=120$nm, $d_{x4}=130$nm, $d_y=15$nm). (b) Comparison of the radiative rate enhancement (over the free-space value) of a dipole embedded in the central gap of $T_3$ (shown as red arrow) and in the gap of an isolated dimer with the same parameters. Out-of-plane (c,e) and in-plane (d,f) intensity distributions (log scale) of the electric field radiated by a dipole located in the gap of the dimer (c,d) and in the central gap of $T_3$ (e,f) at the peak wavelength.

To reveal the differences in the mechanisms of the hot spot formation in the isolated nanodimer and the linear-chain nanostructures shown in Fig. 9a, we compare the vector fields of the on-resonance powerflow through the nanostructures. In Figs. 9i,j,l we show the vector maps of the Poynting vector plotted in the plane cutting through the center of the interparticle gaps of the nanodimer (Fig. 9i), structure $T_1$ (Fig. 9j) and structure $T_2$ (Fig. 9l), respectively. It can be immediately seen that while in the case of the isolated dimer the power lines are simply 'squeezed' into the dimer gap, in $T_1$ the field enhancement results from the formation of a pair of optical vortices of opposite directions (see Fig. 9j). These 'external' nanogears (with the light circulating outside of the nanoparticles) drive the local field enhancement into the dimer gaps. A schematic of the plasmonic two-gear transmission $T_1$, which reverses the light flow through its center, is shown in Fig. 9k. Similarly, the five-dimer structure $T_2$ operates as a four-gear VNT, which generates enhanced forward powerflow through the central nanodimer (Figs. 9l-m).

It should be emphasized that the nanostructures in Fig. 9 have sub-wavelength footprints and reduced scattering losses, which is in stark contrast with the previously-proposed plasmonic arrays designed based on antenna-theory concepts.[21] The narrow resonant features in the scattering response of the external plasmonic VNTs (Fig. 9b) along with the extreme spatial field concentration (Fig. 9g,h) pave the road to engineering VNT-based ultra-sensitive bio(chemical) detectors with high spectral resolution[19, 81, 82]. Also note that different from internal nanogears, the 'size' of an 'external' nanogear – the distance between the phase singularity and the point where the circulating flows of adjacent vortices meet – is not directly determined by the particle size. This opens the way to control and continuously tune the spectral and near-field characteristics of external VNTs.

## Dipole emission shaping with VNTs

For fluorescence sensing or nanoscale quantum information processing, coherent control over interactions between plasmonic nanostructures and dipole sources – as models of quantum oscillators such as molecules and quantum dots (QDs) – is extremely important. Localized emitters coupled to plasmonic nanoantennas can be used as both transmitters and receivers of the optical signal[4, 5, 7, 9, 18]. Typically, plasmonic elements are engineered to enhance optical emission and to direct the optical energy either into free-space radiation[7, 9, 18] or into the propagating surface plasmon-polaritons[83-85]. The latter approach is especially promising as it reduces scattering losses and yields chip-integrated 'dark' plasmonic nanocircuits where all the light-matter interactions occur through the non-radiative near-field coupling mechanism[84, 85]. However, conventional plasmonic waveguides have high dissipative[83, 84, 86-88] and/or scattering[89] losses at optical wavelengths, which challenges construction of extended plasmonic nanocircuitries.

Long dephasing times in combination with nanoscale modal volumes of trapped-mode resonances in VNT structures make them promising platforms for engineering efficient coherent interaction of light with embedded dipole emitters. An example of the plasmonic nanostructure that fulfills this goal is shown in Fig. 10. This structure is engineered to resonantly enhance the radiative rate and to shape the radiation patterns of a single dipole emitter embedded into its center (shown as red arrow in Fig. 10a) by directing the dipole radiation into a sequence of coupled optical vortices. In particular, as can be seen in Fig. 10b, embedding the dipole in the focus of the VNT structure shown in Fig. 10a results in the manifold resonant enhancement of its radiative rate over that of a dipole in the gap of an isolated dimer. The time-averaged Poynting vector corresponding to a free-space oscillating electric dipole with the dipole moment $\mathbf{p}(t) = p_0 \mathbf{u} \cos(\omega t)$ has the following form: $\mathbf{S}(\mathbf{r}) = (3P_0/8\pi r^2)\mathbf{r}\sin\alpha$ ($P_0 = (ck_0^4/12\pi\varepsilon_0)p_0^2$, $\mathbf{u}$ is a unit vector, $\mathbf{r}$ is the unit vector into the direction of observation, $k_0$ is the wave number, $\alpha$ is the angle between vector $\mathbf{u}$ and the observation direction $\mathbf{r}$).

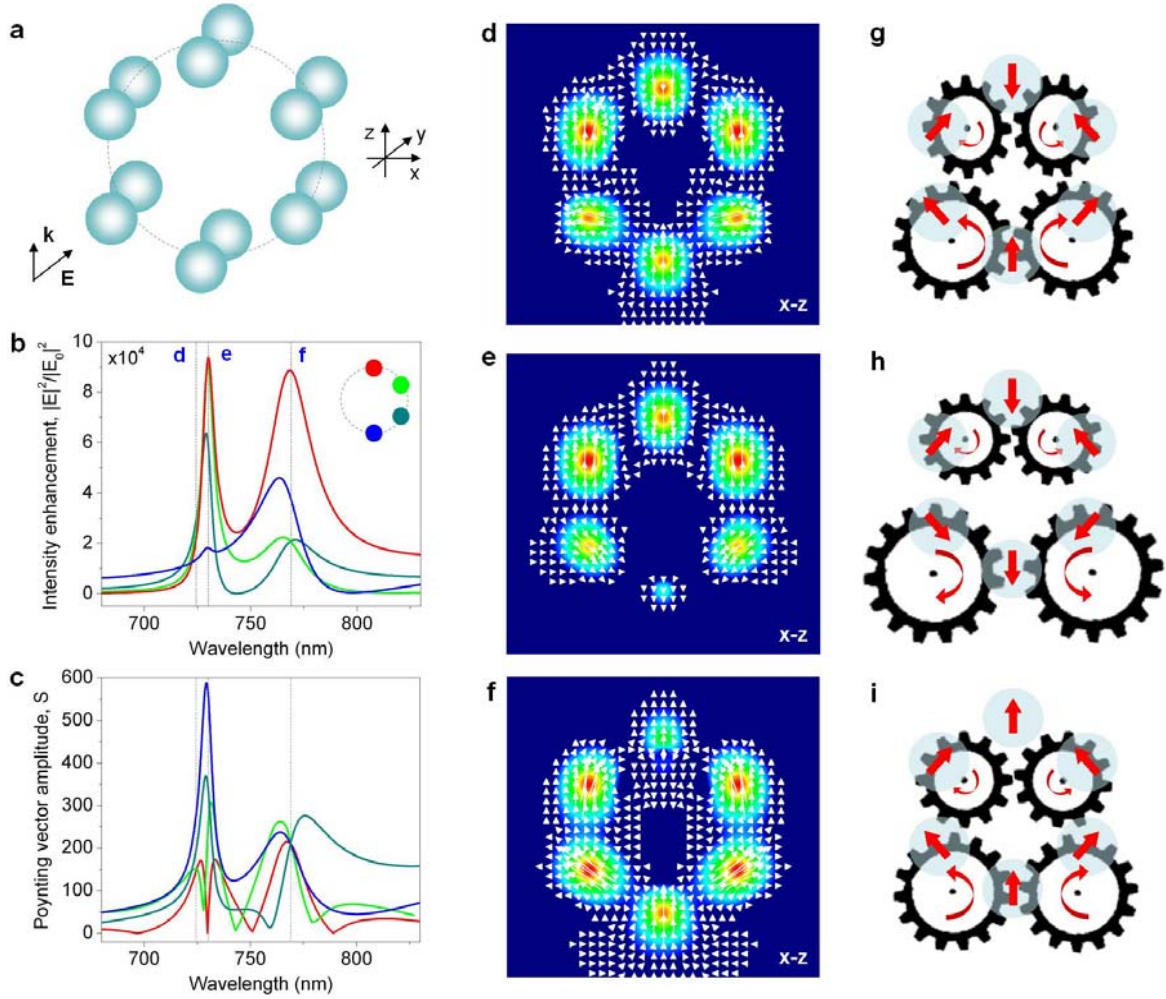

**Fig. 11** Re-configurable looped external nanogear transmission. (a) Schematic of a ring-like arrangement of Ag nanoparticle dimers ($r$=50nm, $d_x$=10nm, $d_y$=3nm, ambient index $n$=1.33). Wavelength spectra of the electric field intensity enhancement (b) and the Poynting vector amplitude (c) in the gaps of four dimers in the ring whose positions are marked in the inset to (b). The colors of the lines correspond to the colors of the position markers. The nanodimer ring is illuminated by a plane wave with the electric field polarized along the dimers axes. (d-f) Poynting vector intensity distribution and powerflow through the nanostructure cross-section at three different wavelengths marked with dashed lines in (b,c). (g-i) Schematics of the coupled and uncoupled VNTs generated by the structure at the three selected wavelengths. Light flux in each nanogear is looped between the particles.

Accordingly, the free-space dipole is characterized by the outward powerflow along straight radially-diverging lines, with no flow along the dipole axis ($\alpha = 0$), and the strongest flow in the directions $\alpha = \pm\pi/2$ [90].

Embedding the emitter into the gap of the nanodimer antenna boosts its radiative rate over that of the free space value (see Fig. 10b), but does not significantly modify the outward powerflow pattern. As Figures 10c and d show, the nanodimer-coupled dipole emits light within a very broad angular range, with preferential emission into $\alpha = \pm\pi/2$ directions. In stark contrast, the optical powerflow from the dipole source embedded in VNT $T_3$ is significantly modified at the resonant wavelength corresponding to the formation of vortex nanogears, resulting in the strong angular shaping of the dipole emission pattern observed in Figs. 10e,f. The emission pattern shaping is determined by the fine structure of the powerflow in the immediate neighborhood of the source, which is imposed by the VNT nanostructure. The field emitted by the VNT-embedded dipole exhibits a vortex structure, with the light being recirculated through the VNT and subsequently leaking at the chain edges (see Figs. 10e,f). It should be noted that efficient re-circulation of dipole emission through the VNT can potentially be useful for enhancing and engineering many-body phenomena between two or more embedded charged particles through the VNT-induced renormalization of their direct Coloumb interaction[91]. The long dephasing times of the VNT resonances should translate into the increase of the interaction range over that observed in conventional low-Q plasmonic nanostructures.

## Adaptive light routing through VNTs

Just as mechanical gears and hydrodynamic turbines form the basis of complex machinery, vortex nanogears can be combined into complex plasmonic networks to enable nanoscale light routing and switching. Although optical vortices are sensitive to changes in the geometrical configuration of the nanostructure, upon small parameter perturbations, they are typically structurally stable and only change their location[49]. This opens the way to continuously tune the VNT characteristics in a controllable fashion by changing the nanostructure design. On the other hand, large perturbations of the controllable external parameters (such as e.g. the wavelength of the incident field) can be introduced to force vortices of opposite sign either to approach and annihilate, or to nucleate as a pair[73]. These effects can form the basis for realizing active spatio-temporal control of light flow on the nanoscale.

The optoplasmonic nanogate shown in Fig. 8 is the simplest example of a reconfigurable VNT structure. One possible realization of a fully-plasmonic reconfigurable VNT is shown in Fig. 11a. It consists of six Ag nanodimers arranged into a symmetrical ring-like structure. The spectra of the electric field intensity in the gaps of the nanodimers comprising this structure are plotted in Fig. 11b. Here, the four lines correspond to the intensity enhancement probed in the centers of four dimers, whose positions are indicated by different colors in the inset. The intensity spectra feature two prominent peaks (Fig. 11b). The spectrum of the Poynting vector amplitude at these positions reveals a frequency-dependent powerflow of alternating directions (Fig. 11c) through the dimer gaps.

The vector fields of the powerflow evaluated at three different wavelengths (marked with dashed lines in Figs. 11b,c) show drastically different pictures of light circulation through the nanostructure. In particular, formation of a looped VNT composed of four coupled vortex nanogears is observed in Figs. 11d,g. In contrast, when the same structure is illuminated by light of a slightly shifted wavelength, it behaves as a pair of uncoupled two-gear transmissions, resulting in the reversal of the local light flow (Figs. 11e,h). At the wavelength of its second resonant peak, the structure once again acts as a coupled four-gear transmission (Figs. 11f,i). The centers of the vortex nanogears are, however, located at different locations than those in Figs. 11d,g, which results in the reversal of the flow direction through the gap of the upper dimer. The possibility of changing the direction of the optical powerflow in the structure shown in Fig. 11a, which has a footprint of ~320 nm$^2$, simply by inverting the rotation of the nanogears opens up tremendous opportunities for light routing and switching on the nanoscale.

## Hydrodynamic analogy

We have shown that the VNT concept can intuitively explain the observed complex electromagnetic interactions in plasmonic nanostructures. The fundamental mechanisms underlying this intuitive picture can be revealed by using the mathematical isomorphism between the hydrodynamic equations and the electromagnetic wave equations separated into phase and amplitude variables[92-97]. In particular, the wave equation for a linearly-polarized monochromatic wave $\mathbf{E}(\mathbf{r},t) = \mathbf{U}(\mathbf{r})\exp\{i(\Phi(\mathbf{r}) - \omega t)\}$ propagating in a piece-wise homogeneous linear nonmagnetic medium with a complex permittivity $\varepsilon(\mathbf{r}) = \varepsilon'(\mathbf{r}) + i\varepsilon''(\mathbf{r})$ can be reduced to the Navier-Stokes-like equations for a special case of the steady flow of 'photon fluid' (PF)$^+$:

$$\nabla(\rho(\mathbf{r})\mathbf{v}(\mathbf{r})) = \alpha(\mathbf{r})\rho(\mathbf{r}), \qquad (1)$$
$$(\mathbf{v}(\mathbf{r}) \cdot \nabla)\mathbf{v}(\mathbf{r}) = -\nabla(V(\mathbf{r}) + Q(\mathbf{r})). \qquad (2)$$

Equations (1) and (2) resemble the continuity (mass conservation) equation and the Euler (momentum conservation) equations of fluid dynamics respectively, where the intensity $I(\mathbf{r}) = |\mathbf{U}(\mathbf{r})|^2$ plays the role of the PF density $\rho(\mathbf{r})$, and the phase gradient plays the role of the fluid velocity, $\mathbf{v} = \nabla\Phi(\mathbf{r})$. Here, $k_0 = \sqrt{\varepsilon_0 \mu_0}\omega$ is the wavenumber, $\alpha(\mathbf{r}) = -k_0^2 \varepsilon''(\mathbf{r})$ describes material loss or gain, and has no direct hydrodynamic analogy, $V(\mathbf{r}) = k_0^2/2(1 - \varepsilon'(\mathbf{r}))$ is the external potential created by the boundary conditions on the nanostructure (in the free space, $V(\mathbf{r}) = 0$), and $Q(\mathbf{r})$ is the internal 'quantum' potential, which has no analog in hydrodynamics. Accordingly, the optical flux defined by the Poynting vector transforms into the analog of a fluid flux (the momentum density): $\mathbf{S} = 1/(2\mu_0\omega) \cdot \rho(\mathbf{r})\mathbf{v}(\mathbf{r})$.

One important difference between conventional fluids and the PF is that the latter has no true conservation of mass since photon 'mass' can be created owing to the linear gain $\varepsilon''(\mathbf{r}) < 0$ and dissipated through material losses $\varepsilon''(\mathbf{r}) > 0$. The 'photon mass' reduction due to dissipative losses necessarily leads to the decrease of the PF density (i.e., field intensity). The hydrodynamic analogy highlights the importance of designing plasmonic nanostructures capable of recirculating light energy outside of the metal (such as external VNTs in Figs. 8-11). Furthermore, the PF represents the special case of a potential flow[92], whose velocity field is described as the gradient of a scalar function (velocity potential) and thus has zero vorticity ($\nabla \times \mathbf{v} = \nabla \times \nabla\Phi \equiv 0$) everywhere except at the centers of the vortices created by the field singularities, which are known as 'free' or 'irrotational' vortices. Unlike real fluids, the PF has quantized properties as the electric field value is invariant with respect to a change of its phase $\Phi(\mathbf{r})$ by an integer multiple of $2\pi$ [94, 96]. Finally, the compressibility of the PF – which is determined by internal and external potentials – is nonlocal, and a conventional local pressure term appears in (2) only if nonlinear self-defocusing effects are included[95, 96]. Despite the unique features of the PF, analogies between 'photonic' and hydrodynamic fluids are extremely useful for the prediction and elucidation of intricate optical effects in plasmonic nanostructures in the context of the VNT approach, as we will show in the following.

In the case of a steady flow of the PF, the field patterns are constant in time, yet a local convective acceleration/deceleration of the flow can occur between different parts of the plasmonic nanostructure. In the structures shown in Figs. 6-11, these local changes of flow rate are driven by the formation of free vortices, with each vortex line inducing a velocity field given by the Biot-Savart formula[98]. In particular, the tangential velocity of the free vortex varies inversely with the distance from its center, and the angular momentum is thus uniform everywhere throughout the vortex-induced circulating flow.

A *change* in the fluid's momentum can generate *pressure*, and this hydrodynamic effect is utilized in hydraulic pumps and motors, which increase the fluid kinetic energy (angular momentum) and then convert it into usable pressure energy. We will now re-consider the problem of wave scattering by the linear VNT $T_2$ shown in Fig. 9a by invoking the fluid dynamic analogy. In Fig. 12a we plot the evolution of the PF velocity and density along the *z*-axis, which passes through the center of the VNT focal point (the velocity vector that points in the direction of the phase change has only *z*-component along this line). At the VNT focal point, the flows generated by adjacent counter-rotating vortices collide and form a 'shock wave' in the form of a region of high PF density. In the situation depicted in Fig. 12a, the PF convectively accelerated by the potential forces (governed by Eq. 2) impacts onto the narrow interparticle gap of the central dimer of $T_2$. The threading of the PF through the $T_2$ gap leads to a sudden change in the flow regime and, according to Eq. 1, results in the dramatic increase of the PF local density driven by the conversion of the PF kinetic energy into pressure energy. In effect, when illuminated by a plane wave, the nanostructure shown in Fig. 9a operates as a plasmonic analog of a hydraulic pump. More specifically, it can be classified as a 'natural-circulation turbopump,' which exploits the convective acceleration of the PF caused by its circulation around the phase singularities 'pinned' to the nanostructure to generate localized areas of high PF density.

The hydrodynamics-electromagnetics analogy can also be used to develop new strategies for the VNT design and performance improvement. For example, the fact that the tangential velocity of the circular flow in the free vortex varies inversely with the distance from its center of rotation (phase singularity) can be exploited to increase the VNT focal point intensity. Fig. 12b demonstrates how variations in the distance between the centers of adjacent counter-rotating vortices comprising the $T_2$ structure from Fig. 9 translate into the electric field intensity enhancement at the VNT focal point. In particular, the field intensity can be boosted by reducing the gaps between neighboring dimers, owing to the increase in the PF kinetic energy at the collision point of the vortices. We emphasize that this effect relies on the gap size reduction in the direction perpendicular to the polarization direction of the incident electric field, and, therefore, cannot be intuitively predicted within the conventional framework of the dipole coupling between closely-located nanoparticles. For larger separations, the linear growth of the PF density at the VNT focal point with the decreased inter-dimer distances (Fig. 12b) reflects the inverse proportionality of the vortex-induced velocity to the nanogears radii. For smaller inter-dimer separations ($d_x < 50nm$) increased near-field plasmonic coupling between neighboring dimers causes deviations from the linear behavior of the PF density. Unlike the flow of hydrodynamic fluids in external potential fields, the PF flow through the VNT is driven by the external potential $V(\omega, \mathbf{r})$ that depends on the wavelength of the incident light. As a result, a change of the nanostructure geometry causes a spectral shift of the resonant feature corresponding to the VNT formation (Fig. 12b).

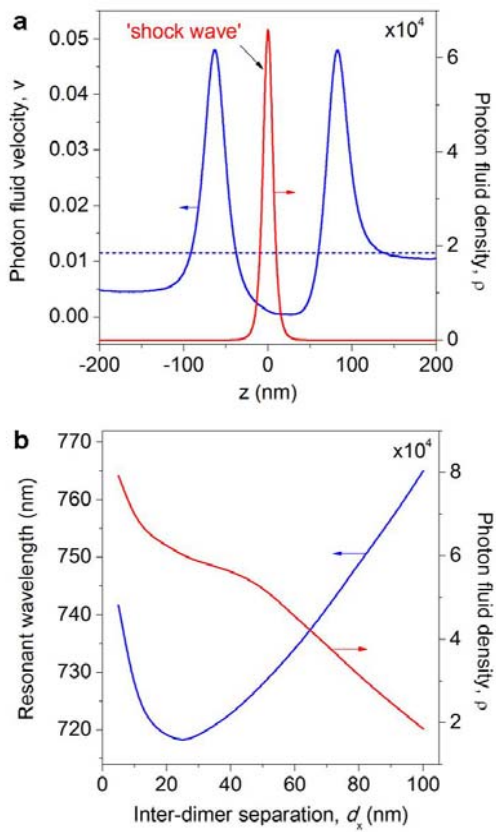

**Fig. 12** VNT as an analog of a hydraulic pump. (a) Velocity (blue) and density (red) of the PF along the line parallel to the incident plane wave propagation direction and passing through the focal point of the VNT $T_2$. The dash blue line corresponds to the constant PF velocity in the absence of the nanostructure. Formation of the 'shock wave' accompanied by the spatially-localized increase of the PF velocity in the VNT central dimer gap is observed. (b) Variation of the PF density (field intensity) (red) in the VNT focal point and a spectral position of the resonant 'shock wave' feature corresponding to the formation of vortex nanogears (blue) as a function of the distance between adjacent dimers comprising the VNT.

## Conclusions

We demonstrated an alternative approach to engineering efficient interactions of light with plasmonic and optoplasmonic nanostructures through electromagnetic theory and numerical simulations that augments conventional antenna theory. Whereas nanoantenna theory analyzes coherent interactions between fields radiated by oscillating nanoparticle dipole moments, the new approach relies on molding the nanoscale powerflow through a series of coupled optical vortices, 'pinned' to rationally-designed nanostructures. In essence, the design focus shifts from achieving *constructive interference* of partial electromagnetic fields to creating points or lines of *destructive interference* (i.e., phase singularities) at pre-designed positions. The structures sustaining coupled-vortex sequences were termed vortex nanogear transmissions, or VNTs. Several examples of both internal (i.e., those that direct the optical powerflow through nanoparticle volumes) and external (i.e., capable of guiding and circulating the optical powerflow through the ambient medium outside of the metal) VNTs were discussed in detail. It was demonstrated that formation of VNTs in plasmonic nanostructures results in a strong increase of Q-factors and corresponding local field intensities. Although the number of discussed applications of the VNT concept was necessarily limited in this manuscript, we emphasize that the VNT enabled engineering of light circulation is a platform technology. The VNT approach provides new opportunities for enhancing nonlinear processes, for designing slow-light components with nanoscale footprints, for efficiently harvesting light energy for photovoltaic and sensor applications, and for designing dynamically-reconfigurable nanocircuits. It is important to note that the VNT approach does not assume different mechanisms of electromagnetic wave interactions with plasmonic nanostructures than Maxwell's equations. Neither does it rely on any approximations of the field distributions, material properties, or boundary conditions. However, it enables analogies to fluid dynamics for describing the light propagation through VNTs, which greatly alleviates an intuitive understanding of the complex electromagnetic interactions that govern the optical responses of these unique plasmonic nanostructures. We believe that these analogies may eventually facilitate a transfer of other hydrodynamic engineering concepts and device designs to photonics and plasmonics. This approach may benefit from the application of efficient unstructured grid solvers derived from the well-developed methods of computational fluid dynamics[99-101] to simulate and design VNT of arbitrary configurations.

While this work focuses on the description of the theoretical concept of VNTs, it should be noted that the discussed structures can be readily fabricated by standard techniques such as electron-beam lithography[25, 27, 102], template-assisted self-assembly[41, 103], nanoassembly,[104] etc. Furthermore, although here we considered only the structures composed of spherical nanoparticles, which are ameable to robust modeling with semi-analytical electromagnetic algorithms, the general approach is by no means restricted to a certain particle shape, size or material composition.

## Acknowledgements

The work was partially supported by the National Institutes of Health through grant 5R01CA138509-02 (BMR), the National Science Foundation through grants CBET-0853798 and CBET-0953121 (BMR) and the Army Research Laboratory Cooperative Agreement W911NF-06-2-0040 (BMR). SVB thanks Anton Desyatnikov (Australian National University) and Jason W. Fleischer (Princeton University) for stimulating discussions, and Daniel W. Mackowski (Auburn University) for making his Fortran codes publicly available.

Notes and references

[a] Department of Chemistry & The Photonics Center, Boston University, Boston, MA 02215, USA; E-mail: sboriskina@gmail.com
[b] Department of Chemistry & The Photonics Center, Boston University, Boston, MA 02215, USA. E-mail: bmr@bu.edu

[+] Application of the Madelung transformation[93, 96, 97] $\mathbf{E}(\mathbf{r}) = \sum_{m=1}^{3} \hat{e}_m U_m(\mathbf{r}) \exp\{i\Phi(\mathbf{r})\}$ to the electric wave equation that arises from Maxwell's equations for a linearly-polarized monochromatic wave in a linear isotropic (piecewise)homogeneous medium in the absence of sources, $\nabla^2 \mathbf{E}(\mathbf{r}) + k_0^2 (\varepsilon'(\mathbf{r}) + i\varepsilon''(\mathbf{r})) \mathbf{E}(\mathbf{r}) = 0$, brings it to the hydrodynamic form, with the intensity playing the role of a 'photon fluid' density $\rho(\mathbf{r}) = \mathbf{U}(\mathbf{r}) \cdot \mathbf{U}(\mathbf{r})$, and the phase gradient the role of the fluid velocity, $\mathbf{v} = \nabla \Phi(\mathbf{r})$. The real part of the transformed wave equation yields the energy conservation law $\mathbf{v}(\mathbf{r}) \cdot \mathbf{v}(\mathbf{r})/2 + V(\mathbf{r}) + Q(\mathbf{r}) = E_0$. The sum of kinetic, external potential $V(\mathbf{r}) = k_0^2/2(1-\varepsilon'(\mathbf{r}))$ ($V(\mathbf{r}) = 0$ in the free space) and internal 'quantum'[93-97] potential $Q(\mathbf{r}) = 1/(8\rho(\mathbf{r})) \sum_{m=1}^{3} (\nabla \rho_m(\mathbf{r}) \cdot \nabla \rho_m(\mathbf{r})/\rho_m(\mathbf{r})) - \nabla^2 \rho(\mathbf{r})/(4\rho(\mathbf{r}))$ energies is invariant and equal to the integration constant $E_0 = k_0^2/2 \neq E_0(\mathbf{r})$, which is the total energy of the system ($\rho_m(\mathbf{r}) = U_m^2(\mathbf{r})$). Applying the Nabla operator to the energy conservation law and taking into account the potential character of the flow, the momentum conservation equation for the steady flow of 'photon fluid' ($\partial \mathbf{v}/\partial t = 0$) can be obtained: $(\mathbf{v}(\mathbf{r}) \cdot \nabla)\mathbf{v}(\mathbf{r}) = -\nabla(V(\mathbf{r}) + Q(\mathbf{r}))$. The imaginary part of the wave equation transforms into the mass-conservation-like equation for the steady flow ($\partial \rho/\partial t = 0$) in the presence of sources and sinks: $\nabla(\rho(\mathbf{r})\mathbf{v}(\mathbf{r})) = \alpha(\mathbf{r})\rho(\mathbf{r})$, $\alpha(\mathbf{r}) = -k_0^2 \varepsilon''(\mathbf{r})$. Hydrodynamic-like (or 'conservation') form of the electromagnetic wave equation can also be obtained in the most general case, including propagation of time-dependent fields in non-linear inhomogeneous media